# Long valley lifetime of dark excitons in single-layer WSe$_2$

Yanhao Tang[1]*, Kin Fai Mak[1,2,3], Jie Shan[1,2,3]

[1] School of Applied and Engineering Physics, Cornell University, Ithaca, NY, USA.
[2] Laboratory of Atomic and Solid State Physics, Cornell University, Ithaca, NY, USA.
[3] Kavli Institute at Cornell for Nanoscale Science, Ithaca, NY, USA.
*Correspondence to: yt378@cornell.edu

**Single-layer transition metal dichalcogenides (TMDs) provide a promising material system to explore the electron's valley degree of freedom as a quantum information carrier[1–3]. The valley degree of freedom in single-layer TMDs can be directly accessed by means of optical excitation[4–6]. The rapid valley relaxation of optically excited electron-hole pairs (excitons)[7–9] through the long-range electron-hole exchange interaction[10,11], however, has been a major roadblock. Theoretically such a valley relaxation does not occur for the recently discovered dark excitons[12–16], suggesting a potential route for long valley lifetimes[10]. Here we investigate the valley dynamics of dark excitons in single-layer WSe$_2$ by time-resolved photoluminescence spectroscopy. We develop a waveguide-based method to enable the detection of the dark exciton emission, which involves spin-forbidden optical transitions with an out-of-plane dipole moment. The valley degree of freedom of dark excitons is accessed through the valley-dependent Zeeman effect under an out-of-plane magnetic field. We find a short valley lifetime for the dark neutral exciton, likely due to the short-range electron-hole exchange[17,18], but long valley lifetimes exceeding several nanoseconds for dark charged excitons.**

Single-layer transition metal dichalcogenides (TMDs, MX$_2$: M=Mo, W; X=S, Se) are direct band-gap semiconductors with direct gaps located at the K and K' valleys of the Brillouin zone[19,20]. Both valence and conduction bands are spin-split at the two valleys by the strong spin-orbit coupling. The exciton formed by Coulomb interaction from electron and hole of antiparallel spins is a bright exciton (optically active), and from electron and hole of parallel spins, a dark exciton (optically inactive). The bright exciton exhibits strong valley circular dichroism (i.e. each handedness of circularly polarized light couples only to one of the two valleys), which provides an effective means to access the valley degree of freedom[4–6]. However, the valley relaxation is very fast (order of 10 ps) for the bright neutral[7–9] and charged excitons[21–23]. The fast valley relaxation is attributed to the long-range electron-hole exchange interaction[10,11], which mixes the two valley exciton states. On the other hand, intervalley scattering of the dark exciton would require a spin flip, which does not occur through the long-range exchange interaction[10]. A long-lived valley polarization of the dark exciton is thus possible. In tungsten-based TMDs the dark exciton has a lower energy than the bright exciton and has recently been shown long-lived[13,24]. Direct measurement of the valley lifetime of the dark exciton, however, remains challenging. This spin-forbidden exciton has an out-of-plane (OP) transition dipole moment[12,14,25], making its detection difficult with conventional far-field optical techniques. In addition, unlike for the bright exciton, there are no valley-dependent optical selection rules for the dark exciton that can be utilized for direct optical access of the valley degree of freedom.

Here we study the valley dynamics of the dark exciton in single-layer WSe$_2$ by time-resolved photoluminescence (PL) spectroscopy. By coupling WSe$_2$ to a GaSe waveguide (Figure 1c), we demonstrate efficient detection of the dark exciton emission using a conventional far-field setup, and effective selection by polarization of emission originated from an OP and an in-plane (IP) transition dipole moment. The approach complements the reported methods for the dark exciton detection[12–16], for instance by applying a large in-plane magnetic field[13,16] or by near-field coupling to surface-plasmon polaritons[12] or an antenna-tip[15], but is more practical for device applications. In addition, we show that the valley-polarized dark charged exciton (Figure 1a,b) can be initiated through scattering of the valley-polarized bright exciton. Under an out-of-plane magnetic field, the valley-polarized dark excitons from the K and K' valleys can be further separated by the Zeeman shift, enabling the measurement of the valley polarization and dynamics.

The schematic of the device geometry is shown in

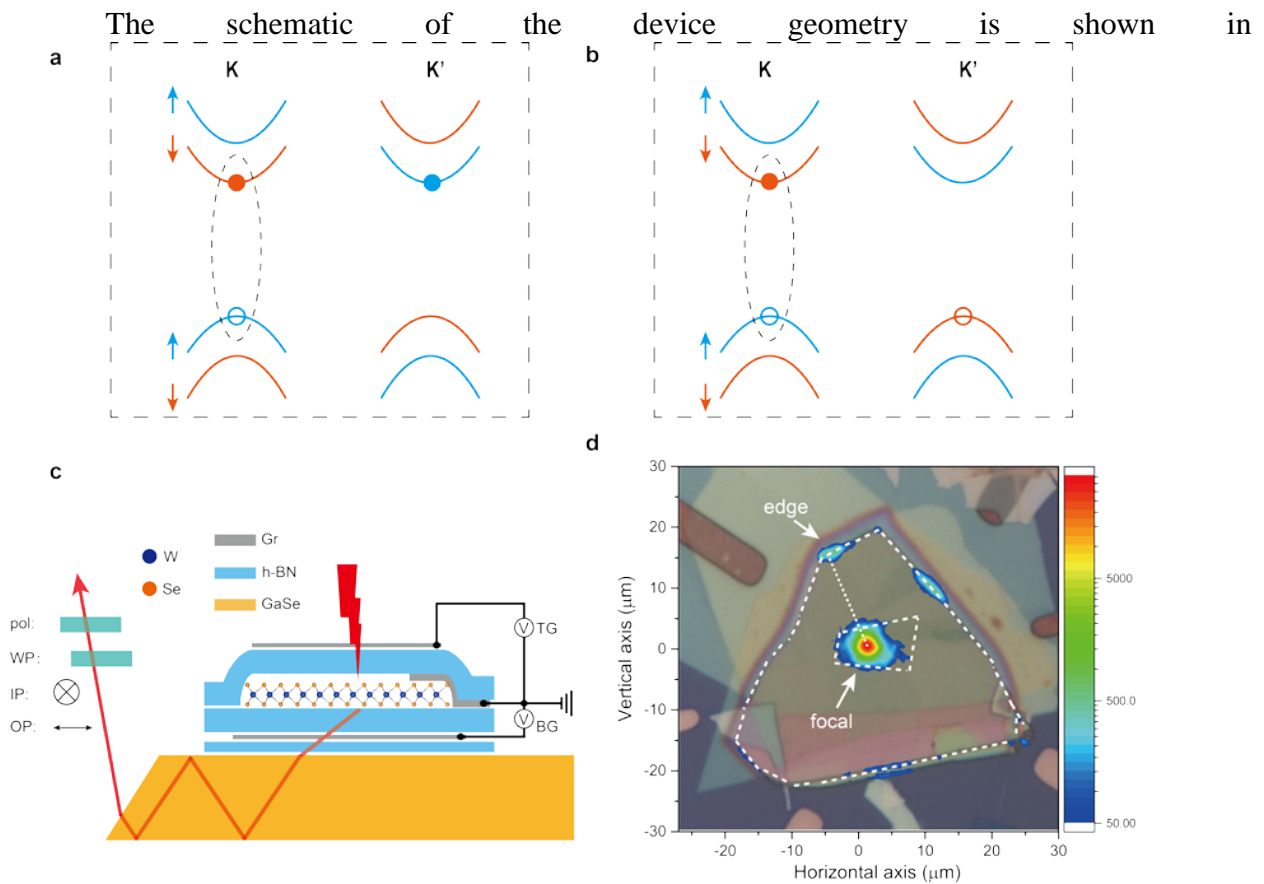

Figure 1c. Single-layer WSe$_2$ is contacted with a few-layer graphite electrode, and can be gated by a top and a bottom graphite gate with hexagonal boron nitride (h-BN) gate dielectrics. The WSe$_2$ field-effect device is positioned on top of a GaSe layer of several-hundred-nanometer thickness, which functions as a slab waveguide[26]. Such a thickness is required to support at least one optical mode. A combination of a half-wave plate and a polarizer selects the emission from an IP or an OP dipole. The entire device is on a Si substrate with an oxide layer. GaSe was chosen for the waveguide since it has a relatively high optical refractive index (about 2.9), low loss in the WSe$_2$ PL spectral range (several cm$^{-1}$)[27], and a van der Waals crystal structure that allows us to build the entire device using the van der Waals heterostructure platform. (See Methods for details on the device fabrication.)

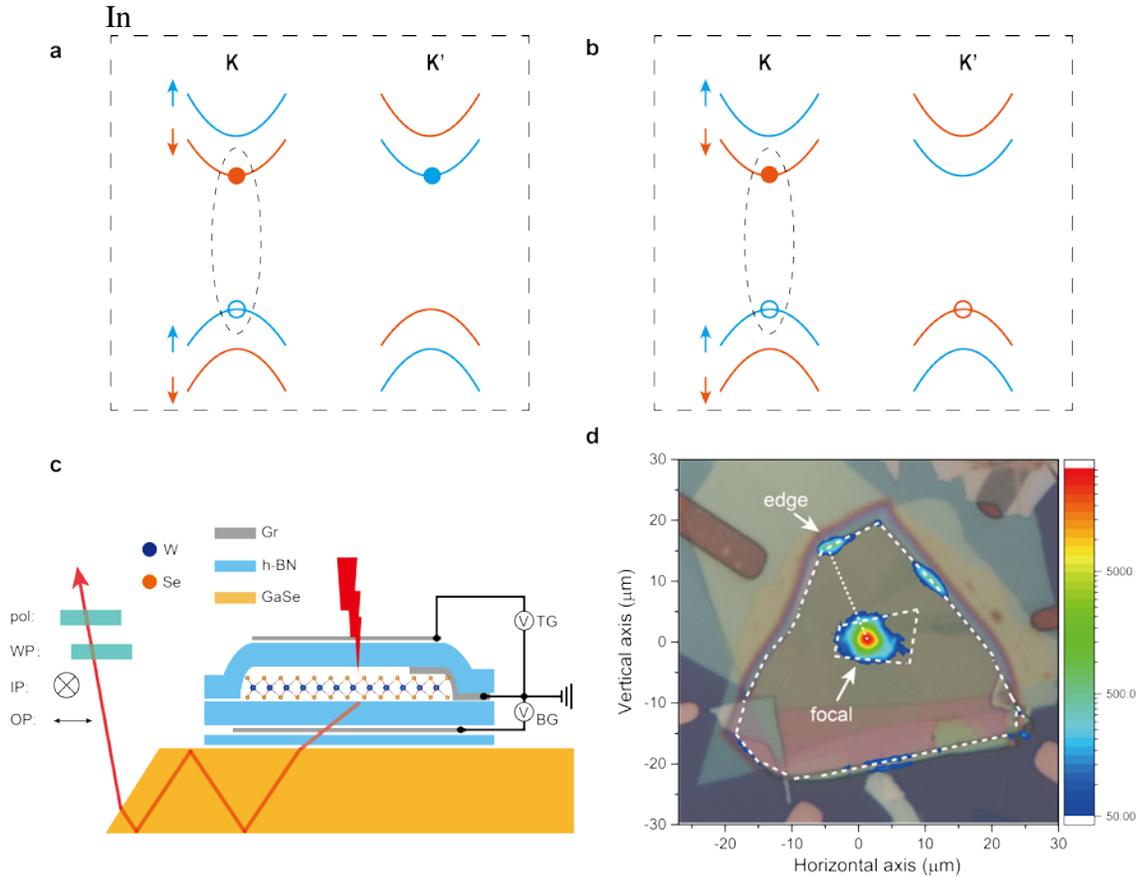

Figure **1**d, an optical image of a typical device is overlaid with its PL image. The PL collected in the back-reflection geometry is observed not only from the focal point, at which the optical excitation is focused, but also from several edges of the waveguide. Light is easier to couple out of the waveguide at those edge faces that are perpendicular to the wave propagation direction (the lines connecting the edge and the focal point are referred to as the focal-edge lines below). The output intensity varies at different locations and as large as 5% of the intensity from the focal point has been observed at individual edges. The value presumably can be improved by optimizing the waveguide such as the tilt angle of the exit face (Fig. 1c). In the current generation of devices we have used GaSe as exfoliated from bulk crystals. Below we present the results from one edge (indicated by an arrow in Fig. 1d). The results from other locations are similar. All measurements were performed at 5 K on single-layer $WSe_2$ with varying doping densities while the electric field perpendicular to the layer was kept approximately at zero (The two symmetric gates were set to the same voltage). (See Methods for details on the PL measurements.)

Figure 2a shows the emission spectrum collected from the edge as a function of polarization direction (vertical axis). $WSe_2$ is hole doped in this example (gate voltages were -2.2 V). All sharp spectral features show a two-fold symmetry and can be divided into two groups with orthogonal polarizations. They exhibit maximum intensities when the polarizer transmission axis is set either perpendicular or parallel to the focal-edge line (labeled IP and OP, respectively, in Fig. 2a). The IP channel is dominated by two features, $X^{0,B}$ and $X^{+,B}$, at the high-energy end and multiple sharp features on a broad background at the low-energy end. They correspond to the bright neutral exciton, the bright positively charged exciton (i.e. hole trion), and the localized or finite-momentum bright excitons in single-layer WSe$_2$[28–30]. The OP channel is dominated by two new features, $X^{0,D}$ and $X^{+,D}$, which are assigned as the dark neutral exciton and the dark hole trion, respectively, according to the literature[12,13]. Since the bright and dark excitons are known to be IP and OP dipoles[12–16,31], this result shows that our device geometry can selectively detect the IP and OP dipole emission by polarization. In Figure 2b we compare the PL spectra collected from the edge and from the focal point. The IP channel is nearly identical to the (rescaled) spectrum from the focal point, further supporting that the IP channel is dominated by the IP dipole emission. The red shift of the bright exciton PL collected from the edge originates from re-absorption of the PL by WSe$_2$ during its propagation in the waveguide. Figure 2c and 2d show the doping dependences of the PL from the two channels. Both positively and negatively charged excitons (bright or dark) can be accessed in a single device by electrostatic gating. We note that the OP channel intensity is at least 20% of the IP channel intensity, but the extinction of the IP dipole emission in the OP channel is not perfect. Bright exciton emission is also visible in the OP channel (Fig. 2d). But it can be subtracted by using the scaled IP channel as background and has negligible impact on the analysis of the dark exciton dynamics below.

To resolve the valley degree of freedom of the dark exciton, we lift the valley degeneracy by the Zeeman effect under an out-of-plane magnetic field.

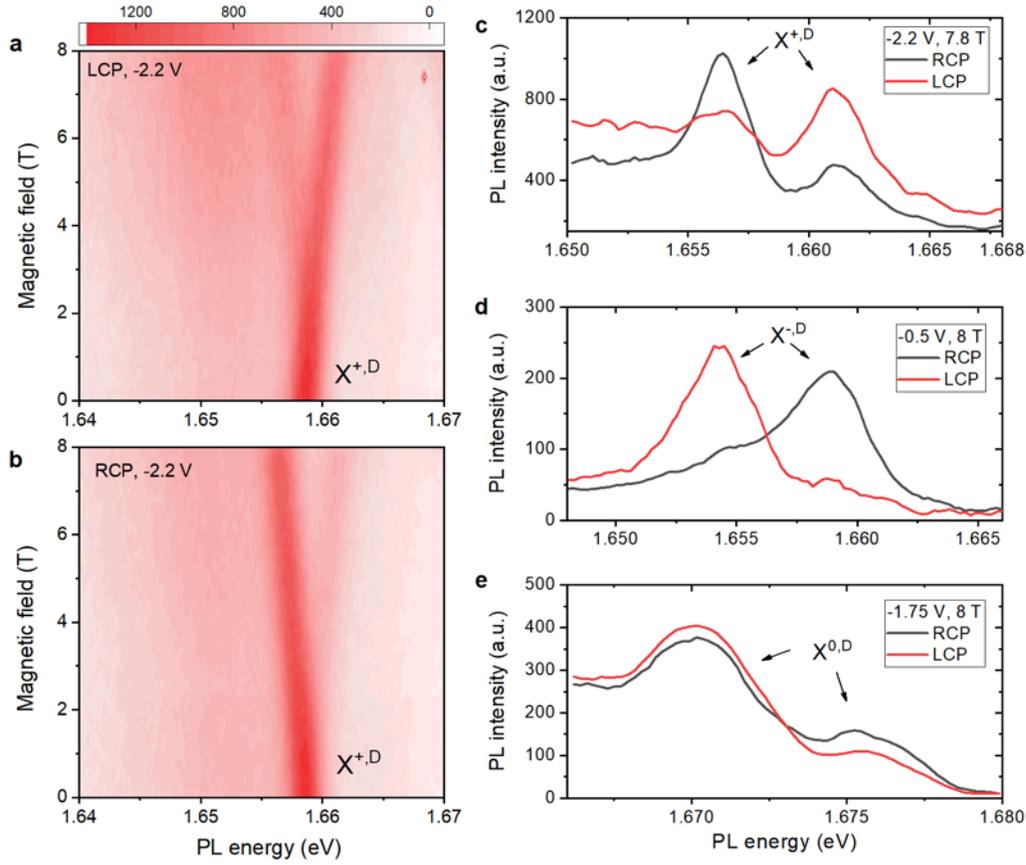

Figure **3**a is the PL spectrum from the OP channel as a function of magnetic field (vertical axis). Single-layer WSe$_2$ is hole doped (both gates at -2.2 V) and is excited by left-circularly polarized (LCP) light at 1.96 eV (well above the bright exciton fundamental resonance). We observe that the dark hole trion PL $X^{+,D}$ splits into two peaks as the magnetic field increases, and the higher-energy peak is brighter than the lower-energy peak. The behavior

of $X^{+,D}$ is similar under right-circularly polarized (RCP) excitation

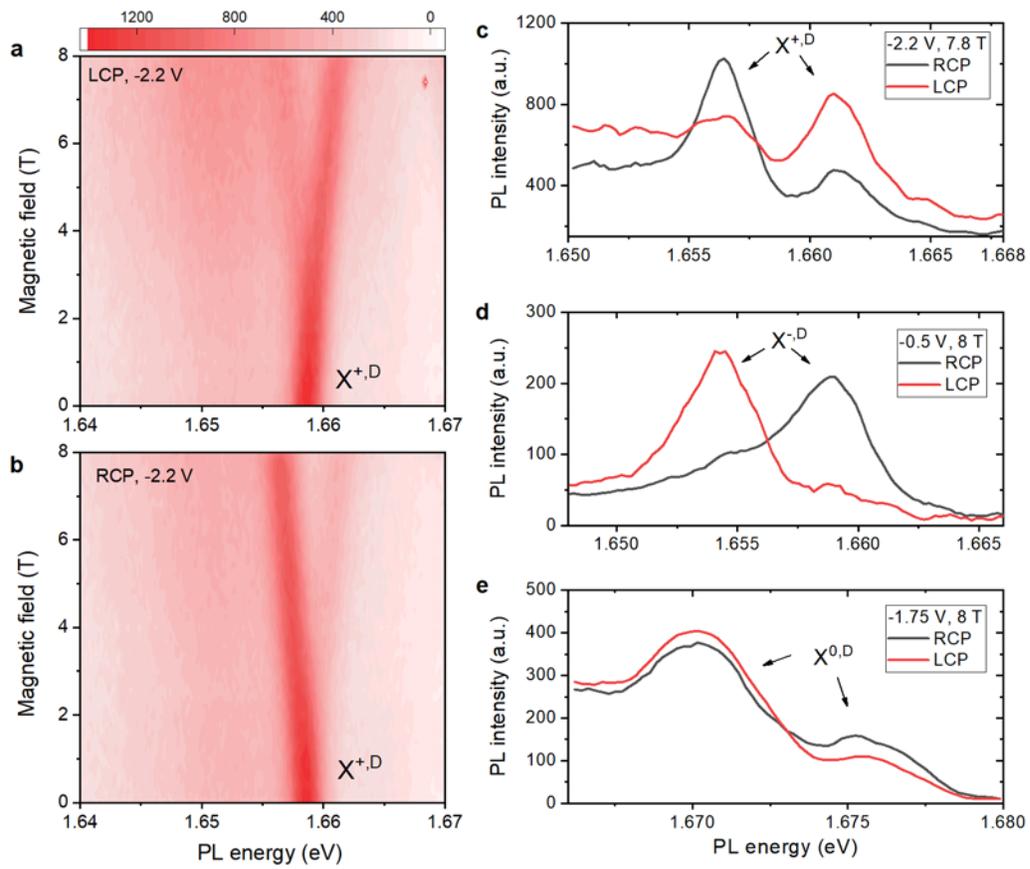

(

Figure 3b), but the brightness of the two peaks is reversed.

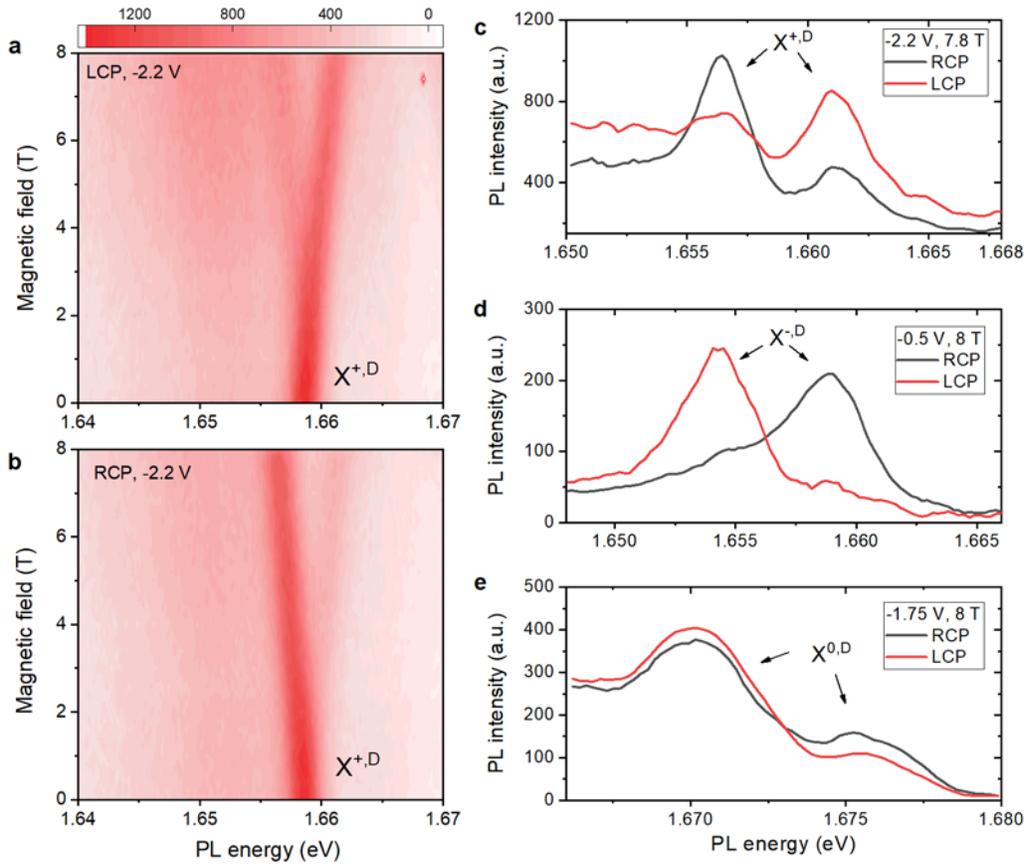

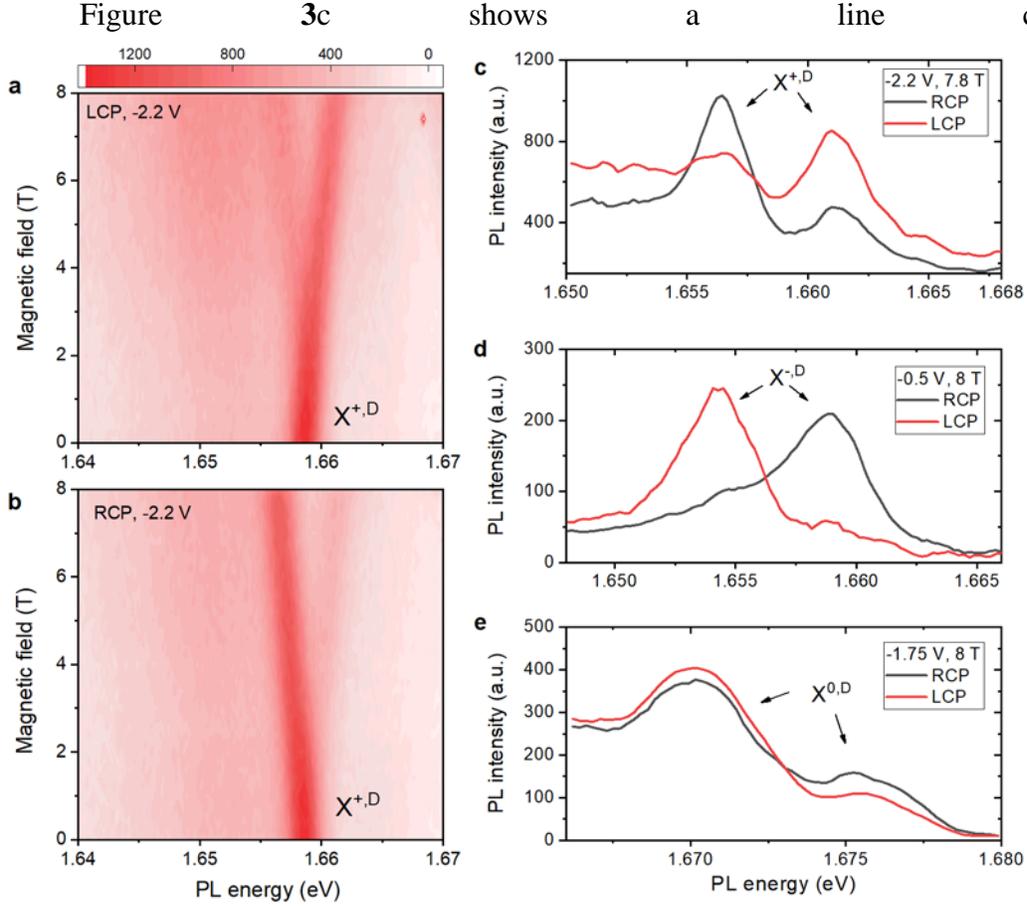

Figure 3c shows a line cut of Figure 3a and 3b at 7.8 T. The observed peak splitting corresponds to a g-factor of $10.9 \pm 0.1$, which is consistent with earlier studies for dark trions[13]. The fact that either of the Zeeman-split states can have a higher intensity than the other under LCP/RCP excitation evidences that the valley polarization can be optically initialized (through the bright exciton relaxation) and maintained in the dark hole trion. The microscopic scattering process of the bright to dark exciton states is not well understood and warrants future studies. By making reference to the behavior under the same experiential conditions of the bright hole trion (for which the valley-dependent optical selection rules apply, see Supplementary Sect. 1), we determine that the higher- and lower-energy Zeeman-split states of $X^{+,D}$ correspond to the K and K' valleys, respectively. The valley contrast can be estimated as $\rho \approx \frac{I_{K'} - I_K}{I_{K'} + I_K}$, where $I_{K(K')}$ is the integrated intensity of the Zeeman-split K(K') valley states. The PL intensity is a reasonable approximation of the valley population when the Zeeman splitting is insignificant. The steady-state value of $|\rho|$ for the dark hole trion is found to be about 40%. A similar steady-state valley contrast is estimated for the dark electron trion $X^{-,D}$ (Fig. 3d), but a negligible valley contrast is observed for the dark neutral exciton $X^{0,D}$ (Fig. 3e). The latter suggests a rapid valley relaxation within the neutral dark exciton lifetime (about 100 ps [24]). Below we focus only on the dark trions ($X^{+,D}$ and $X^{-,D}$), for which the valley lifetime is potentially long. (See Supplementary Sect. 1 for detailed analysis of the g-factor, the valley index for the Zeeman-split states, and the steady-state valley contrast.)

We perform the energy-resolved time-correlated single-photon counting (TCSPC) measurements on doped $WSe_2$ under an OP magnetic field of 8 T. The sample was excited by circularly polarized optical pulses of 180 fs in duration, 79 MHz in repetition rate, and peaked at 1.82 eV. The PL was collected from the edge in the OP channel. The setup has a temporal resolution of 36 ps (the full-width-half-maximum of the instrument response function). (See Methods for more details on the TCSPC measurements.) Figure 4a shows the time-resolved PL at the peaks of $X^{+,D}$ from the K and K' valleys under RCP excitation. The traces have been deconvoluted with the instrument response function and filtered (smoothed) to remove high-frequency noise above 3 GHz, which arises mainly from the numerical deconvolution process. A small number of counts is present before the arrival of the excitation pulse, which is likely the contribution of long-lived localized excitons (the broad background on which $X^{+,D}$ sits in Figure 2b) excited by previous optical pulses. To avoid the influence on our analysis of such background, we have restricted our window of interest (3.5 ns) so that the count at any time is at least twice of the background value. For the RCP excitation, the PL intensity of the K' valley trion is about twice of the intensity of the K valley trion. For the LCP excitation, the intensity trend of the K and K' valleys is reversed (Figure 4b). This is consistent with the steady-state PL measurements

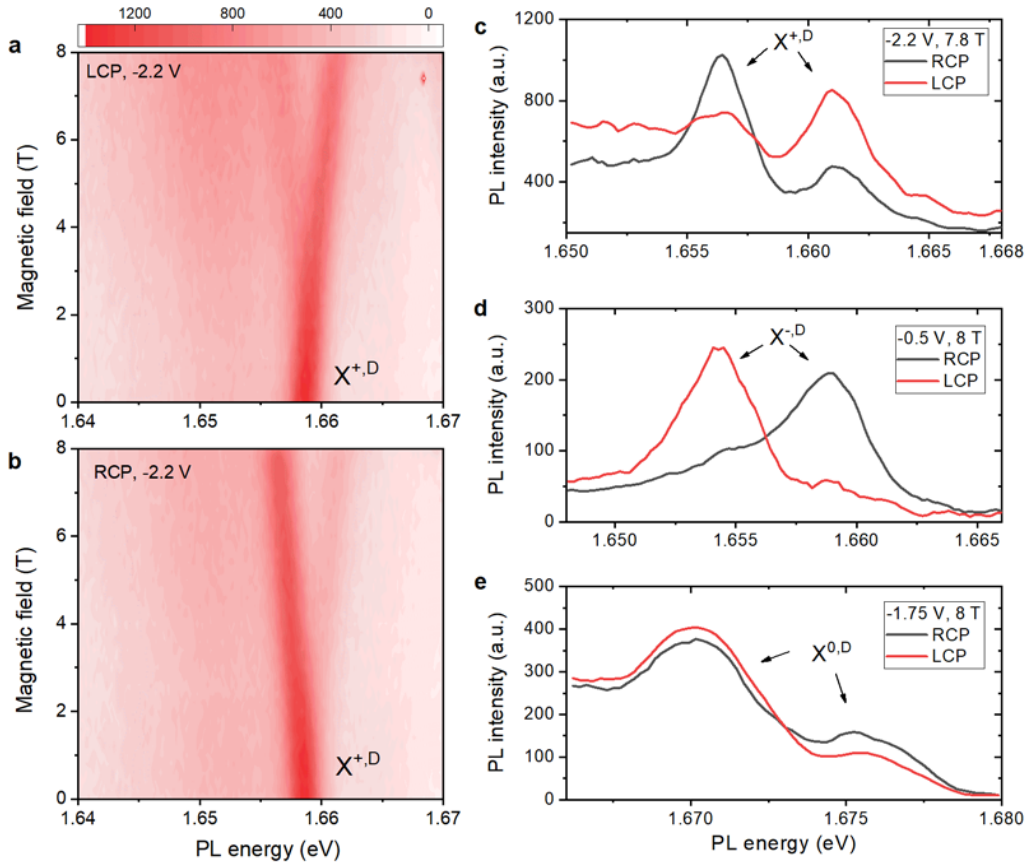

(Figure 3). We evaluate the valley contrast $|\rho(t)|$ using the PL intensities and obtain a decay time constant of $32 \pm 4$ ns and $4.1 \pm 0.2$ ns, respectively, for the RCP and LCP excitation by fitting the data with a single exponential decay function (dotted lines). The longer valley lifetime obtained for the RCP excitation is expected since the Zeeman effect breaks the degeneracy of the K and K' states and scattering from the lower-energy state to the higher-energy state is suppressed. The two values thus provide an upper and lower bound estimate for the valley lifetime. Similarly, we obtain the valley lifetime of the dark electron trion $X^{-,D}$ to be at least $3.5 \pm 0.5$ ns. (Raw data and detailed analysis of the valley lifetime measurements are included in Supplementary Sect. 3 and 4).

We have observed a long-lived few-nanosecond valley polarization for the dark charged excitons, but not for the dark neutral exciton. The initial theory has argued that the intervalley scattering of the dark neutral exciton that requires a spin flip could not occur through the long-range electron-hole exchange (the lowest-order exchange). However, more recent theoretical works have predicted that the short-range electron-hole exchange (a second-order exchange)[17,18] could mix the two dark neutral exciton states at the K and K' valleys to form two new states with a small splitting. The lower-energy dark neutral exciton is a truly dark state, which is both spin and electric-dipole forbidden. The higher-energy state is a nearly dark state, which is spin forbidden, but dipole allowed with an OP dipole. Recent magneto-luminescence experiments have reported a zero-field splitting of 0.6 meV between these states in single-layer $WSe_2$[24]. In our experiment the observed dark neutral exciton is the spin-forbidden exciton with an OP dipole. The observed short valley lifetime could therefore be attributed to the short-range electron-hole exchange. On the other hand, the dark charged exciton is composed of a dark exciton in one valley and a hole (or an electron) in the other valley (Fig. 1a and 1b). The dark charged excitons are not prone to intervalley scattering through the electron-hole exchange because they have non-zero but opposite momentum at the two valleys.

In conclusion, we have developed a waveguide-based method for resolving the dipole orientation in layered materials, which is challenging with conventional far-field optical techniques. By integrating single-layer $WSe_2$ dual-gate field-effect devices directly into the waveguide, we have been able to time-resolve and valley-resolve the emission from the dark exciton states under an OP magnetic field. We have determined a valley polarization lifetime exceeding several nanoseconds for the dark charged excitons. The dark charged excitons with long valley lifetimes in TMDs provide, in addition to the resident carriers[32–36] and the interlayer excitons[37], a potentially useful carrier for information storage and processing based on the valley degree of freedom.

**Methods**
**Device fabrication.** The waveguide-coupled dual-gate $WSe_2$ devices were built from exfoliated van der Waals materials using a layer-by-layer dry transfer method[38]. Atomically thin hBN, graphite and $WSe_2$ flakes were exfoliated from their bulk crystals onto silicon substrates, which were pretreated with ozone plasma. The thickness of single-layer $WSe_2$ flakes was estimated by their optical contrast and confirmed by the PL spectra. The hBN flakes of similar thickness were used as the gate dielectric for both the top and bottom gates. GaSe was exfoliated onto polydimethylsiloxane (PDMS) to obtain hundred-nm-thick slabs of large size for the waveguide. The stamp used for the transfer was made of polypropylene-carbonate-coated PDMS covered by a polycarbonate (PC) layer. In the transfer process, a GaSe slab was first picked up from PDMS and released onto a 300-nm-$SiO_2$/Si chip with pre-patterned gold electrodes at around 180° C. The residual of the stamp on GaSe was removed by dissolving the whole chip in chloroform for a few minutes followed by a rinse in isopropyl alcohol. Other atomically thin flakes were assembled layer by layer first and the entire stack was then released onto the GaSe waveguide. The stamp residual was removed using the same procedure before the optical measurements.

**Photoluminescence measurements at low temperature.** Devices were mounted in a close-cycle cryostat (a Montana or an Attocube system). For steady-state photoluminescence (PL) measurements, a continuous-wave (CW) laser at 633 nm with power less than 100 μW was used to excite $WSe_2$. For time-resolved PL measurements, output from a Ti:sapphire oscillator (Coherent, Chameleon Ultra II) with a repetition rate of 79 MHz, a photon energy centered at 1.818 eV and an average power of 110 μW was used to excite $WSe_2$. The excitation beam was focused to a beam radius of 1 μm on the $WSe_2$ samples using a microscope objective of a numerical aperture of 0.6 or 0.8. The PL was collected by the same objective in the back-reflection geometry, and focused onto an entrance slit of a monochromator (Princeton Instruments, HRS-300MS) with a 600 grs/mm grating. One exit port of the monochromator is connected to a liquid-nitrogen cooled CCD camera (Princeton Instruments, PYL-400BRX) for steady-state PL measurements. The second exit port is coupled to a single-photon detector (SPD from Picoquant, PD-050-CTD) with a telescope, which reduces the size of the image from the exit port to the SPD by a factor of two to increase the PL collection efficiency. The output of the SPD is registered in Picoharp300 for time-correlated single-photon counting (TCSPC) measurements. The setup has a temporal resolution of 36 ps, as determined from the full-width-half-maximum of the instrumental response function.

**Acknowledgments**
This work was supported by the U.S. Department of Energy, Office of Science, Basic Energy Sciences, under Award #DE-SC0019481.

**Author Contributions**
Y.T., J.S., and K.F.M. conceived the project. Y.T. fabricated the device, carried out the measurements, and analyzed the data. All authors wrote the paper.

**Competing Financial Interests**
The authors declare that they have no competing financial interests.

**Data availability**
The data that support the plots within this paper and other findings of this study are available from the corresponding author upon reasonable request.

**Figure Legends:**

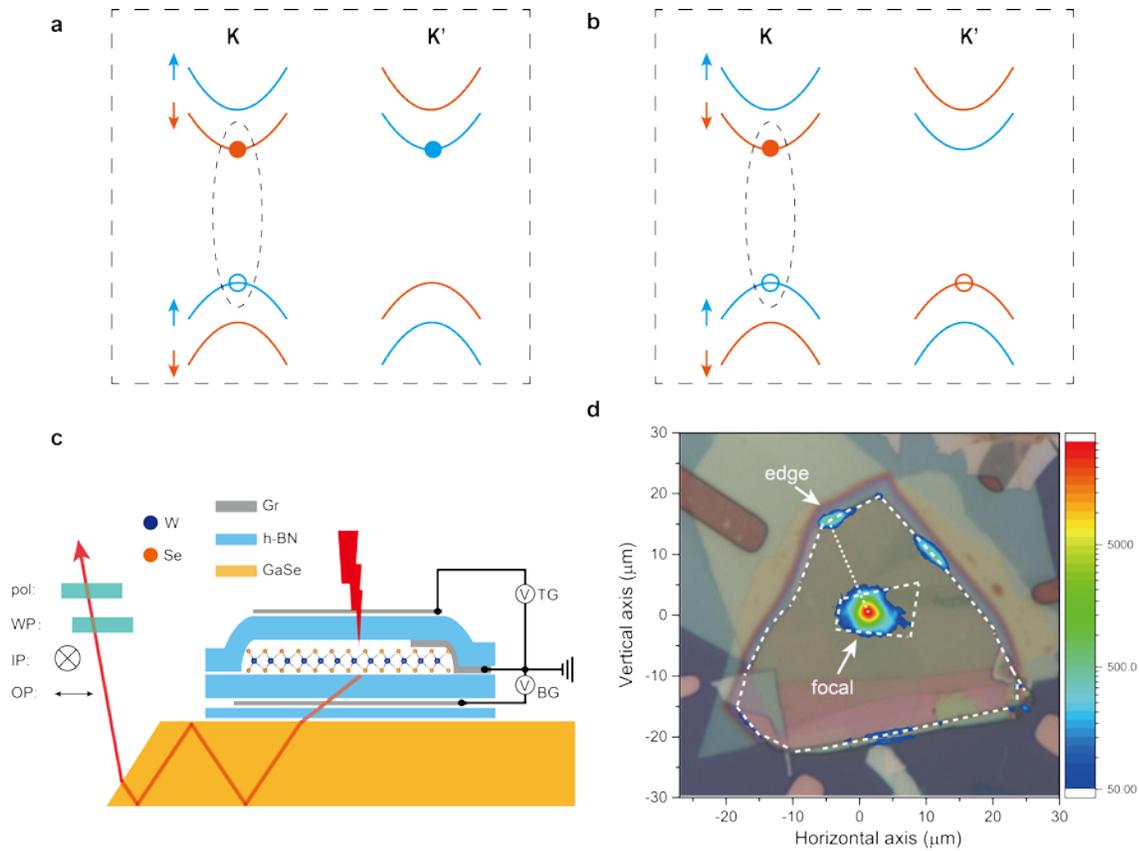

**Figure 1 | Excitons in WSe$_2$ and experimental geometry. a**, **b**, Electronic configuration of a dark negatively charged exciton (**a**) and dark positively charged exciton (**b**) in single-layer WSe$_2$. Blue and orange curves represent electronic bands with electron spin up and spin down, respectively. The hole spin is opposite to what's shown for the electron spin. Dashed ellipses indicate the electron-hole pairs involved in the recombination. **c**, Schematic side view of a dual-gate WSe$_2$ device on a GaSe waveguide. WSe$_2$ is excited by a focused light beam (red lightning symbol) and the resultant PL is guided by the waveguide (red arrowed line). The IP and OP emission dipoles are selected by a half-wave plate (WP) and a polarizer (pol). WSe$_2$ is grounded. TG and BG are the top and bottom gate voltages, respectively. **d**, Optical reflectance and PL images (overlaid) of a typical device. Inner and outer white dashed lines show the boundary of WSe$_2$ and GaSe, respectively. The dotted white line, which is perpendicular to the edge, is referred to as the focal-edge line. The color bar represents the PL intensity.

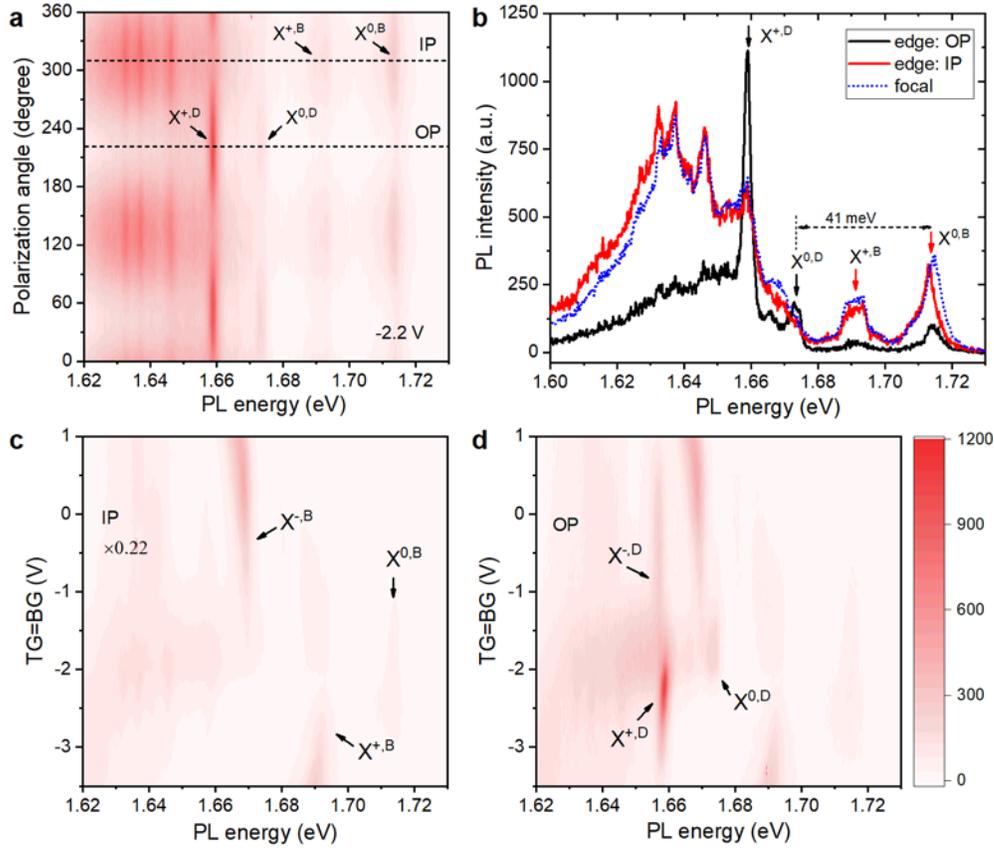

**Figure 2 | Resolving IP and OP dipoles by polarization of the waveguide output**. **a**, Contour plot of the edge PL spectrum as a function of polarization direction for a hole-doped $WSe_2$ (both gates at - 2.2 V). Black dashed lines indicate the polarization corresponding to the OP and IP channels. **b**, Comparison of the PL spectrum from the edge OP (black line) and IP (red line) channels and from the focal point (blue dotted line). The latter is rescaled to match the edge IP channel spectrum. **c, d**, Contour plot of the edge PL spectrum as a function of gate voltage for the IP (**c**) and OP (**d**) channels. The IP channel is rescaled by a factor of 0.22 so that the PL intensity of the bright electron trion $X^{-,B}$ in two channels have a comparable intensity. The two gates are set to the same voltage, which varies only the doping density in $WSe_2$ with symmetric top and bottom gates. The color bar represents the PL intensity in **a, c, d**. $X^{0,B}$, $X^{+,B}$, $X^{0,D}$, $X^{+,D}$ and $X^{-,D}$ denote the bright exciton, bright hole trion, dark exciton, dark hole trion and dark electron trion, respectively. The energy splitting between the bright and dark exciton (41 meV) in **b** agrees with the literature value[12,14].

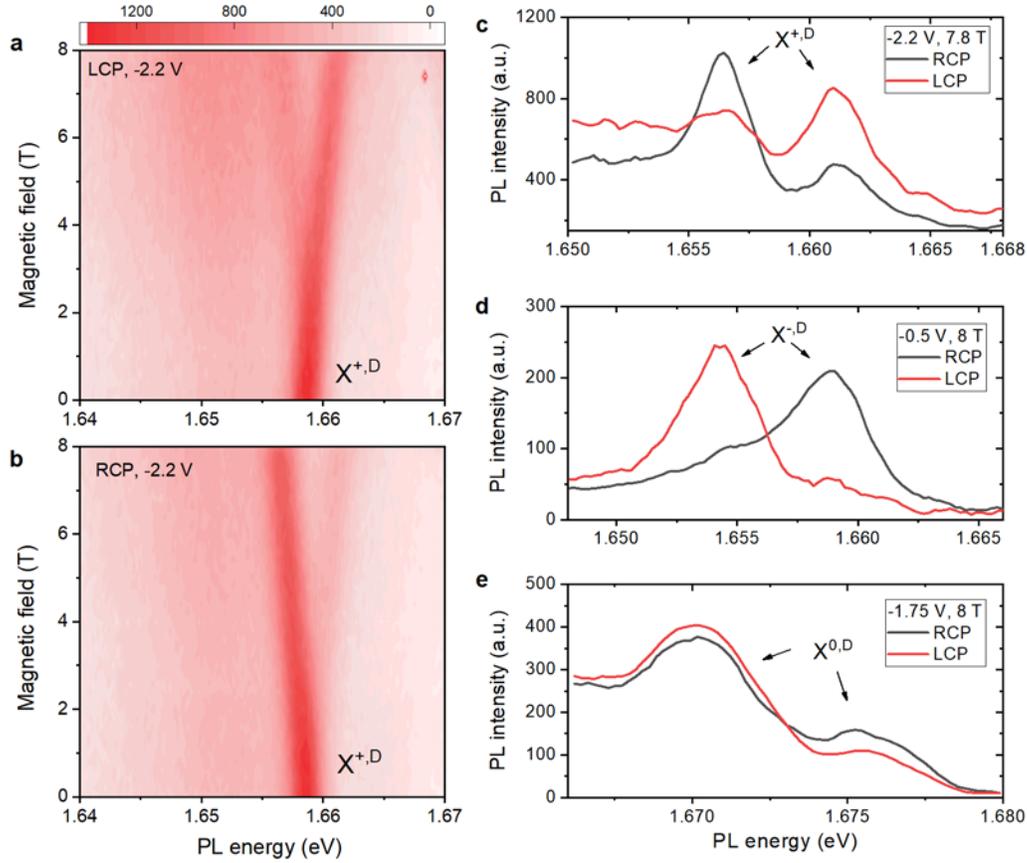

**Figure 3 | Resolving the valley degree of freedom of dark excitons by the Zeeman shift. a**, **b**, Contour plot of the PL spectrum of the OP channel as a function of magnetic field for a hole doped WSe$_2$ (both gates at - 2.2 V). **a** is for the LCP excitation and **b** for the RCP excitation. The color bar represents the PL intensity. **c-e**, PL spectra of the OP channel under RCP (black line) and LCP (red line) excitation. The out-of-plane field is about 8 T. The gate voltages -2.2 V (**c**), -0.5 V (**d**), and -1.75 V (**e**) correspond to a hole-doped, an electron-doped, and a neutral sample, respectively.

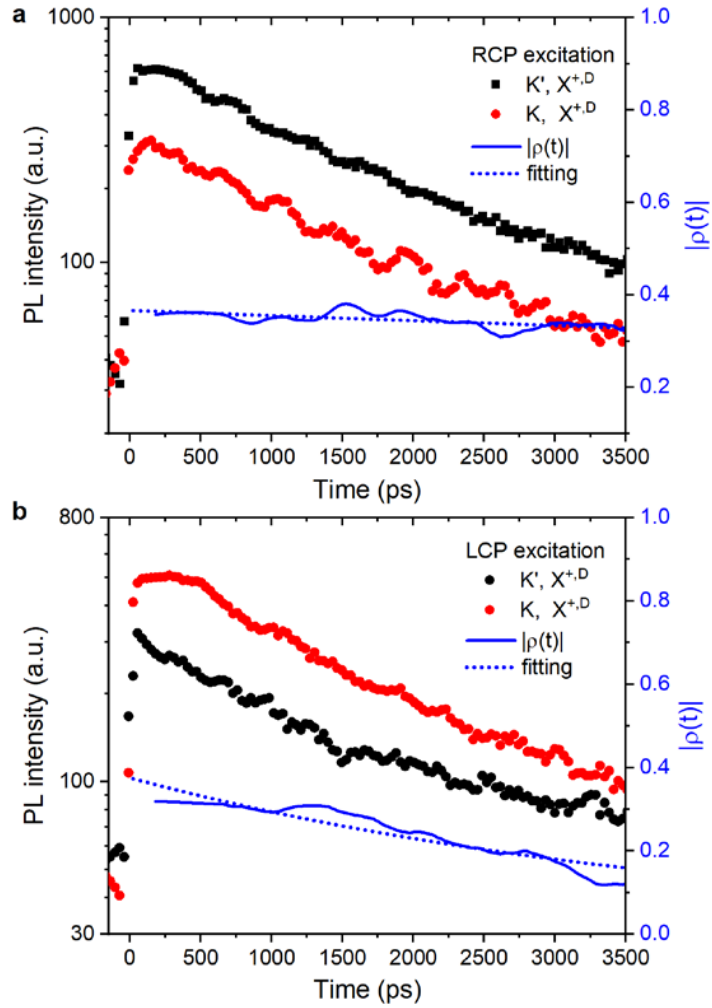

**Figure 4 | Dark exciton valley dynamics.** Time-resolved PL of the dark hole trion in hole doped WSe$_2$ (both gates at -2.2 V) under a magnetic field of 8 T for the RCP (**a**) and LCP (**b**) excitation. Black and red symbols represent the PL of the Zeeman-split dark hole trion associated with the K' and K valleys, respectively. The solid blue curves are the valley contrast $|\rho(t)|$ as defined in the main text. The dotted blue curves are a single-exponential fit, revealing a decay time constant of $32 \pm 4$ ns and $4.1 \pm 0.2$ ns, respectively, for the RCP and LCP excitation.

# Supplementary information for
# Long valley lifetime of dark excitons in single-layer WSe$_2$

Yanhao Tang[1]*, Kin Fai Mak[1,2,3], Jie Shan[1,2,3]

[1] School of Applied and Engineering Physics, Cornell University, Ithaca, NY, USA.
[2] Laboratory of Atomic and Solid State Physics, Cornell University, Ithaca, NY, USA.
[3] Kavli Institute at Cornell for Nanoscale Science, Ithaca, NY, USA.
*Correspondence to: yt378@cornell.edu

1. **Steady-state PL for charged and neutral dark excitons under a magnetic field (Figure S1-S6)**
2. **PL dynamics of excitons (Figure S7, S8)**
3. **Alternative analysis of the valley dynamics (Figure S9)**
4. **Valley dynamics of the dark electron trion (Figure S10)**

1. **Steady-state PL for charged and neutral dark excitons under a magnetic field**

The photoluminescence (PL) spectrum of the out-of-plane (OP) channel was measured for both neutral and doped $WSe_2$ as a function of OP magnetic field. Both the left- and right-circularly polarized (LCP and RCP) excitation above the bright exciton fundamental resonance has been used. We fit each PL spectrum using a superposition of two Gaussians and a smooth background (a polynomial). An example is provided in Figure S4 for the dark hole trion PL under 8 T. A substantial Zeeman splitting between the peak energy of the two Gaussians is observed for the dark hole trion, electron trion, and neutral exciton in Figure S1 – S3, respectively. The g factor is determined to be $10.9 \pm 0.1$, $10.2 \pm 0.3$, and $12.1 \pm 0.7$ accordingly. We can also estimate the valley contrast $\rho \approx \frac{I_{K'} - I_K}{I_{K'} + I_K}$ from the integrated intensity of the two Gaussians $I_{K(K')}$. The valley contrast is about 40% for the dark hole trion measured with both the RCP and LCP excitation (Fig. S4). A similar analysis has been carried out for the dark electron trion (Figure S5) and a valley contrast of about 40% and 60% is obtained using the RCP and LCP excitation, respectively. The uncertainty in this case is much larger due to the lower overall fitting quality and the larger uncertainty in determining the background.

In order to assign the valley degree of freedom for the dark excitons, we measure the PL of the bright excitons through the in-plane (IP) channel under the same experimental conditions and rely on the valley optical selection rules that apply to the bright excitons. One example is shown in Fig. S6 for an electron-doped $WSe_2$ (both gates at 0 V) under 8 T. We observe that the LCP and RCP excitation (that couple exclusively to the K and K' valleys) address the *lower-* and *higher-energy* Zeeman-split states of the bright electron trion, respectively. This is the same order for $X^{-,D}$ and $X^{0,D}$ shown in Figure 3 of the main text. Accordingly, we assign the brighter Zeeman-split peak of the dark excitons with the LCP and RCP excitation to the K and K' valleys, respectively.

2. **PL dynamics of excitons**

The time-resolved PL at different energies of both the IP and OP channels has been measured under zero magnetic field using the time-correlated single-photon counting method. Figure S7 shows the time-resolved PL monitored at the peak energy of the bright exciton $X^{0,B}$ and the bright trion $X^{+,B}$ together with the instrument response function (IRF). They show a fast rise and a slower decay. The decay of the bright exciton and trion is only slightly slower than the IRF. The results from the OP and IP channels are also nearly identical except the PL intensity (not shown). On the other hand, the dark trion shows a substantially longer decay time. Figure S8 shows the time-resolved PL at the dark hole trion energy from the IP and OP channels. The traces have been deconvoluted with the measured IRF. The two channels differ significantly for the first 3.5 ns, with a much slower decay for the OP channel, and become similar gradually afterwards. The latter can be attributed to the contribution of the localized excitons with very long population lifetimes. The localized excitons presumably do not have a well-defined emission dipole direction. The slow decay in the first 3.5 ns for the OP channel corresponds to a single-exponential decay, which yields a population lifetime of $1.60 \pm 0.02$ ns for the dark hole trion.

## 3. Alternative analysis of the valley dynamics

In the main text (Figure 4), we have analyzed the valley dynamics of the dark hole trion by comparing the time-resolved PL at the energy of the K and K' valleys for a given circularly polarized excitation. Alternatively, we can analyze the valley dynamics by comparing the time-resolved PL at a given valley under the RCP and LCP excitation. An example is shown in Fig. S9. A similar valley lifetime is obtained as in the main text. We note that the valley dynamics monitored at the valley with the lower energy shows a faster decay. This asymmetry can be qualitatively explained that the scattering from the higher-energy state to the lower-energy state is energetically favored.

## 4. Valley dynamics of the dark electron trion

Figure S10 shows the time-resolved PL at the lower-energy Zeeman-split state of the dark electron trion for the LCP and RCP excitation. Each dependence can be fitted with a single exponential decay (not shown) with a time constant of $410 \pm 10$ ps. This population lifetime is shorter than that for the dark hole trion. On the other hand, the valley lifetime can be extracted by fitting the valley contrast dynamics with a single exponential decay. The obtained valley lifetime of $3.5 \pm 0.5$ ns is similar to that of the dark hole trion.

**Figures**

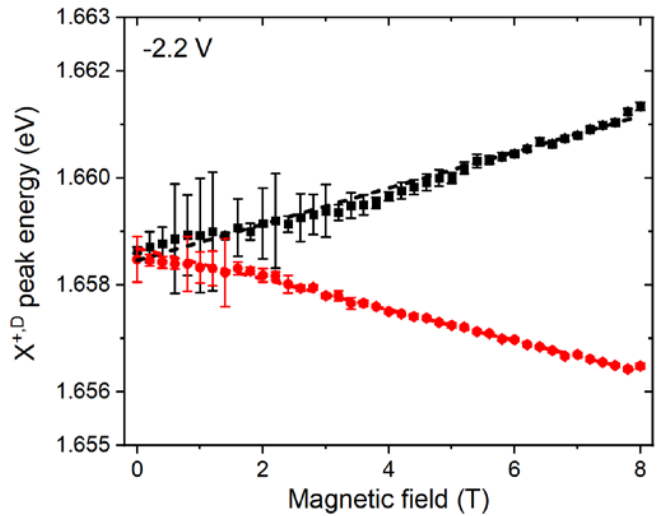

**Figure S1 | Zeeman splitting of the dark hole trion $X^{+,D}$.** The symbols represent the peak energies of the Zeeman-split dark hole trion PL as a function of OP magnetic field. The energies were determined by fitting the PL spectrum using a combination of two Gaussian functions and a smooth background. The error bars are the fitting uncertainty. The dashed lines are linear fits to the data, corresponding to a g factor of $\mathbf{10.9 \pm 0.1}$.

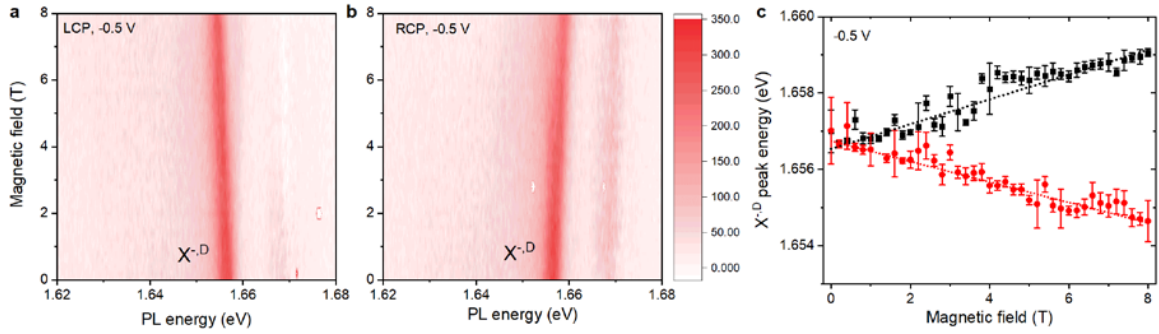

**Figure S2 | Zeeman splitting of the dark electron trion $X^{-,D}$.** **a**, **b**, Contour plot of the PL spectrum of the OP channel as a function of magnetic field for an electron-doped WSe$_2$ (both gates at - 0.5 V). **a** is for the LCP excitation and **b** is for the RCP excitation. **c**, Same as in Fig. S1 for the dark electron trion $X^{-,D}$ which determines the g factor to be $\mathbf{10.2 \pm 0.3}$.

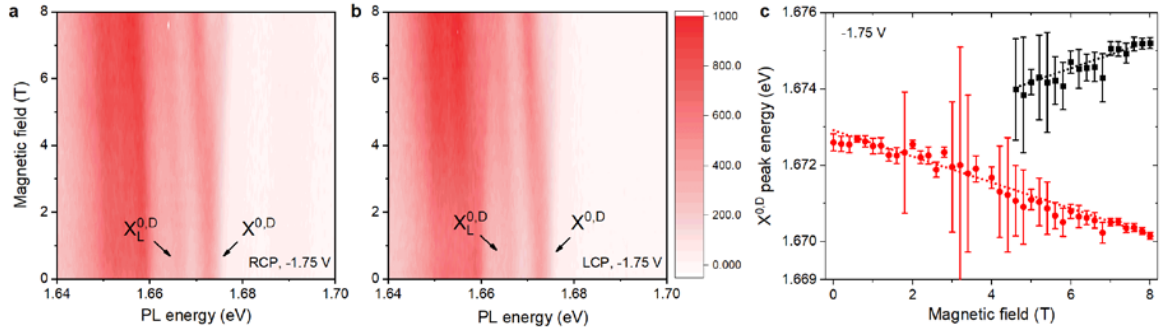

**Figure S3 | Zeeman splitting of the dark exciton $X^{0,D}$.** Same as in Fig. S2 for a neutral WSe$_2$ (both gates at -1.75 V), which determines the g factor for the neutral dark exciton to be $\mathbf{12.1 \pm 0.7}$.

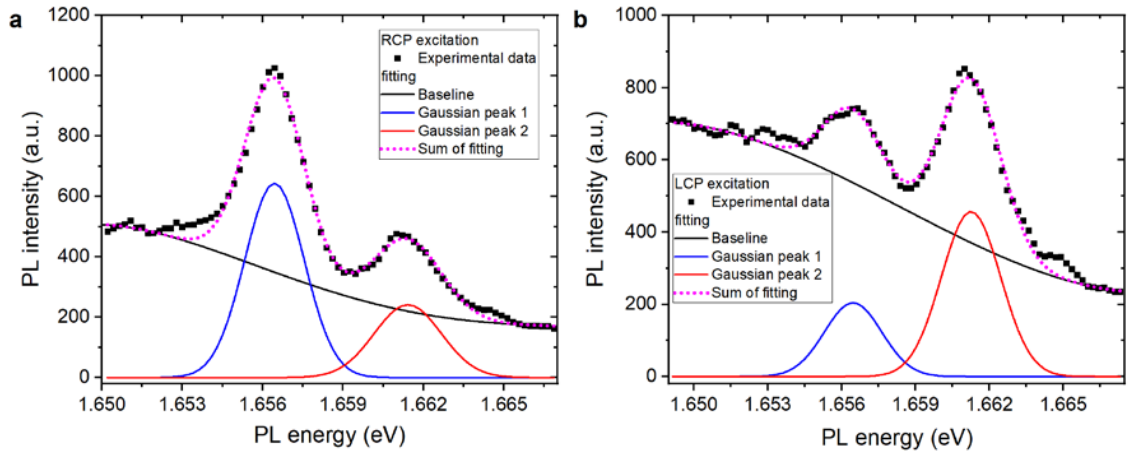

**Figure S4 | Analysis of the PL spectrum under a magnetic field (hole-doped sample).** A hole-doped WSe$_2$ sample (both gates at – 2.2 V) is excited by the RCP (**a**) and LCP (**b**) excitation under 8 T. Each PL spectrum (black symbols) is decomposed into two Gaussians

(blue and red lines) and a smooth background (black line). The dotted magenta curves are the sum of all contributions. The valley contrast, estimated from the integrated emission of the two peaks $A_1$ and $A_2$ as $\left|\frac{A_1-A_2}{A_1+A_2}\right|$, is about 0.4 for both **a** and **b**.

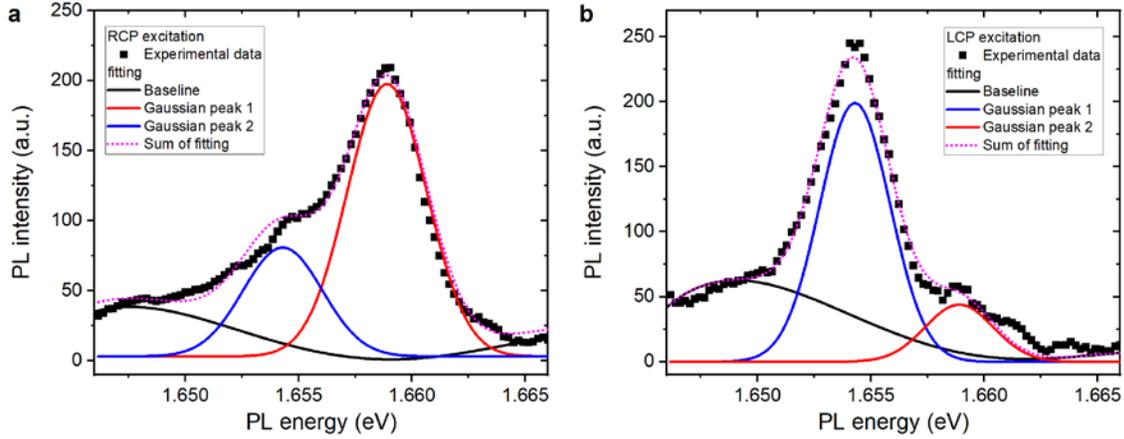

**Figure S5 | Analysis of the PL spectrum under a magnetic field (electron-doped sample).** Same as Fig. S4 but for an electron-doped $WSe_2$ (both gates at - 0.5 V). The valley contrast is estimated to be about 0.4 for **a** and 0.6 for **b**.

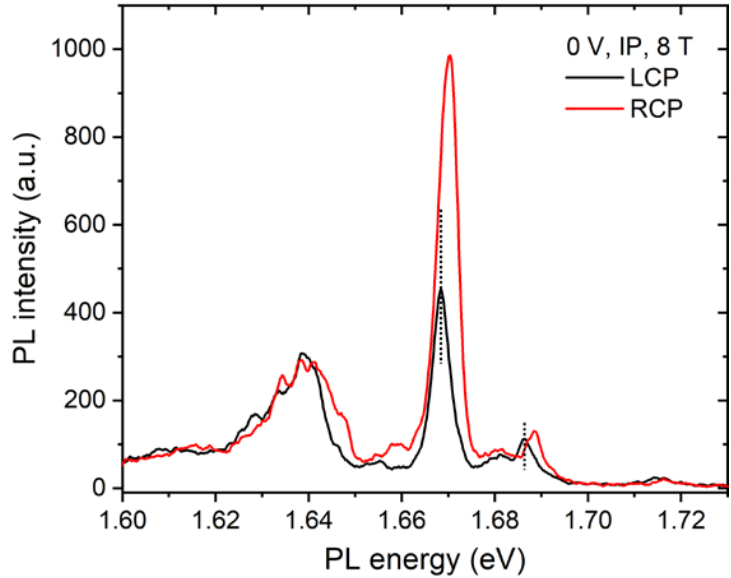

**Figure S6 | PL spectra of the IP channel under a magnetic field.** The PL spectrum of the IP channel for an electron-doped $WSe_2$ (both gates at 0 V) by the LCP (black) and RCP (red) excitation under 8 T. The dotted vertical lines are guides to the eye for the bright electron trion emission peaks under the LCP excitation.

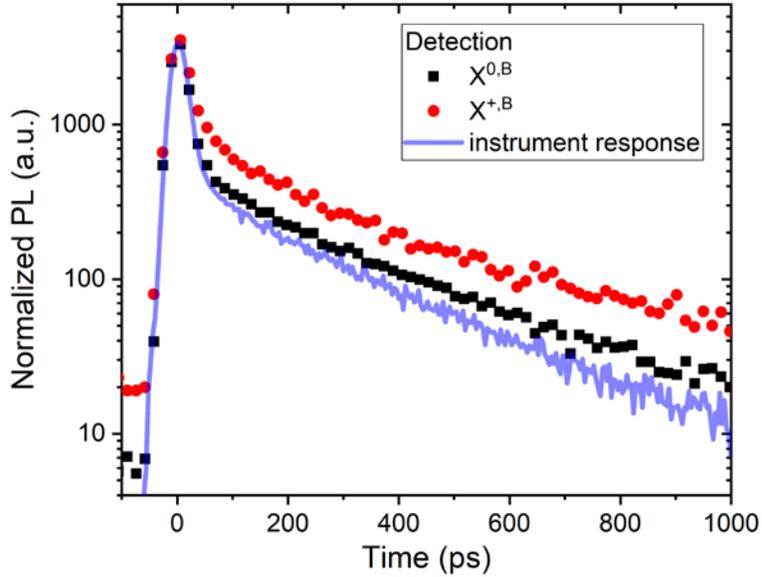

**Figure S7 | PL dynamics of the bright excitons.** Raw data of the time-resolved PL of the bright neutral exciton (black symbols) and hole trion (red symbols) of a hole-doped $WSe_2$ (gate voltages at – 2.2 V). The blue curve is the instrument response function (IRF) obtained by measuring a fs laser pulse.

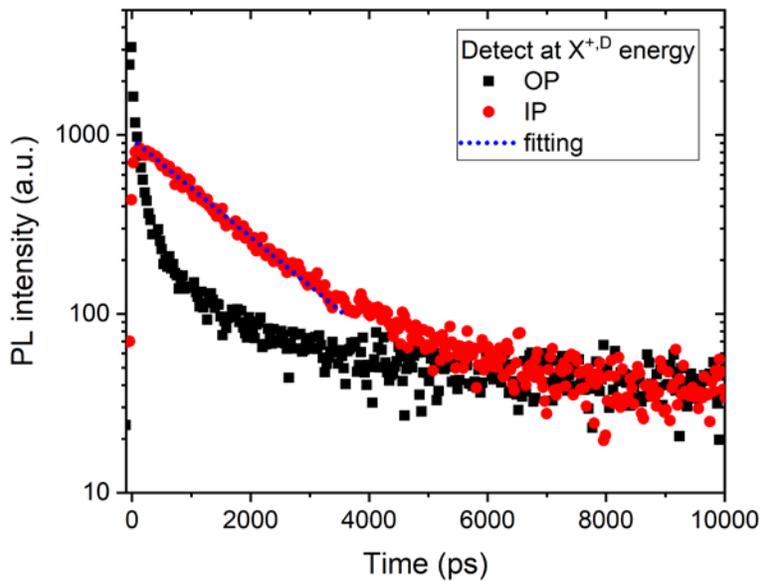

**Figure S8 | PL dynamics of the dark trion.** The time-resolved PL at the energy of the dark hole trion in the OP (black symbols) and IP (red symbols) channel of a hole-doped $WSe_2$ (gate voltages at – 2.2 V). The data have been deconvoluted with the IRF shown in Fig. S7. The blue dotted curve is a single-exponential fit to the OP channel result for the first 3.5 ns. It corresponds to a decay constant of **$1.60 \pm 0.02$** ns.

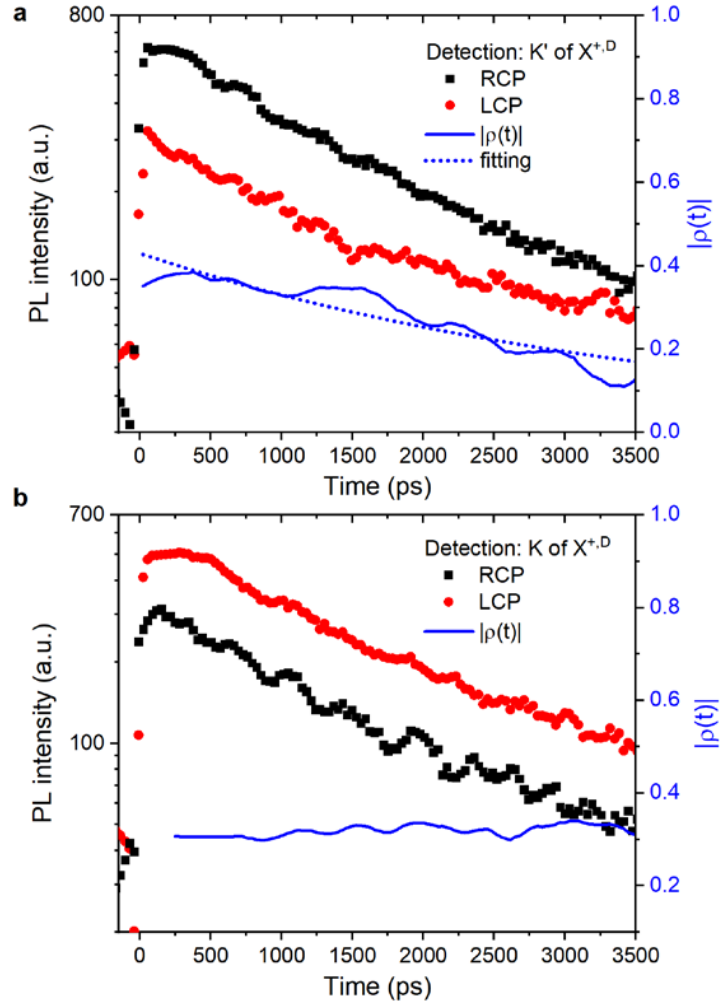

**Figure S9 | PL dynamics of the dark hole trion under 8 T.** The time-resolved PL of a hole-doped $WSe_2$ (both gates at – 2.2 V) is monitored at the K' valley (**a**) and the K valley (**b**) for the dark hole trion. The black and red symbols represent the PL under the RCP and LCP excitation, respectively. The solid blue curves represent the degree of circular polarization (DoCP) defined as $\frac{I_{RCP}-I_{LCP}}{I_{RCP}+I_{LCP}}$, where $I_{RCP(LCP)}$ is the PL intensity with the RCP(LCP) excitation. The dotted blue curve in **a** represents a single-exponential fit to the DoCP dynamics for the first 3.5 ns, corresponding to a decay time constant of $\mathbf{3.8 \pm 0.2}$ ns. No meaningful fit for the DoCP dynamics in **b** can be obtained.

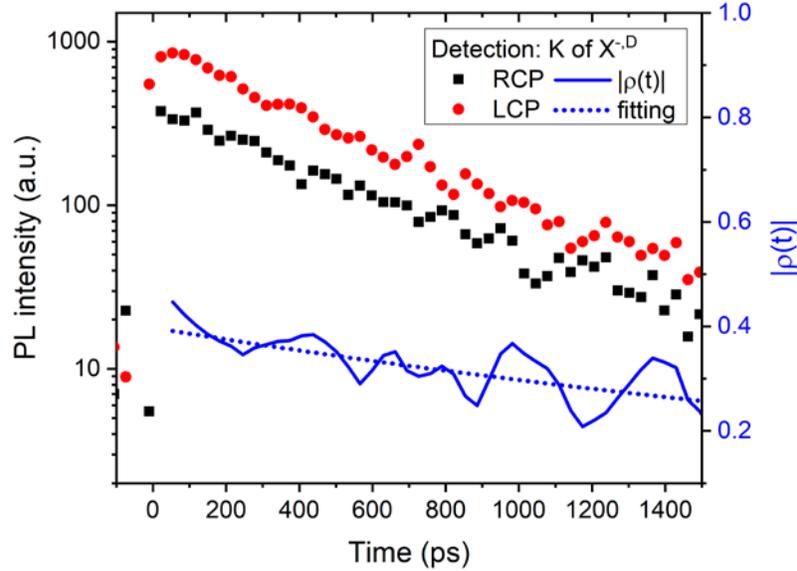

**Figure S10 | Valley dynamics of the dark electron trion.** The time-resolved PL monitored at the K valley of the dark electron trion (the lower-energy peak of the Zeeman-split states) under a magnetic field of 8 T of an electron-doped WSe$_2$ (both gates at - 0.5 V). The black and red symbols represent the PL with the RCP and LCP excitation, respectively. The solid blue curve represents the time-dependence of the degree of circular polarization, and the dotted blue curve is a single exponential fit with a decay time constant of $\mathbf{3.5 \pm 0.5}$ ns for the first 1.5 ns.